\documentstyle[preprint,aps,floats]{revtex}

\begin{document}
\setlength\baselineskip{20pt}

\preprint{\tighten \vbox{\hbox{CALT-68-2212}
		\hbox{hep-ph/9902311}}}

\title{Semileptonic $B$ Decays as a Probe of New Physics}

\author{Walter D. Goldberger\footnote{walter@theory.caltech.edu}\\[4pt]}
\address{\tighten California Institute of Technology, Pasadena, CA 91125}

\maketitle

{\tighten
\begin{abstract}
Physics beyond the Standard Model could be measured indirectly, through its effects on Standard Model observables.  One place to look for such effects is in the semileptonic decays of $B$ mesons.  In order to constrain the possible role of new physics on semileptonic $B$ decays, we present the most general low energy effective Lagrangian constructed from all dimension six four-fermion operators that contribute to the process $b\rightarrow c\ell{\bar\nu}_\ell$.   We then use it to compute the corrections due to new physics to the differential decay rates for the exclusive processes ${\bar B}\rightarrow D^{(*)}\ell{\bar\nu}_\ell$, as well as the inclusive decays ${\bar B}\rightarrow X_c\ell{\bar\nu}_\ell$.  Both inclusive and exclusive rates are expressed in terms of a set of parameters that characterize the types of four-fermion interactions that can be induced by physics beyond the Standard Model.  Although it is not particularly useful to carry out a full analysis until data from the next generation of $B$ factories becomes available, here we illustrate how the existing experimental results can be used to constrain some of these parameters.
\end{abstract}}
\vspace{0.7 in}
\narrowtext

\newpage

\section{Introduction}

Despite the fact that the Standard Model successfully accounts for most of the observed experimental results (see~\cite{hewett} for a review), it is incomplete, and must be replaced by a more fundamental theory at some high energy scale.  Whatever form physics at a higher scale might take, it will generate low energy effective couplings that contribute to Standard Model processes.  These couplings provide the opportunity to search for signatures of new physics indirectly, by measurements of the deviation of experimental observables from their Standard Model predictions.  

Rather than compute the effect of physics beyond the Standard Model within the framework of a specific model, our ignorance dictates that we include all effective interactions consistent with the symmetries of the Standard Model.  The resulting observables are expressed in terms of the parameters of the generalized interaction, which can be constrained by the experimental data.  Since these parameters characterize the most general interaction, their experimental bounds can be used to rule out or guide in the construction of specific models.

Anticipating the precision data from the upcoming generation of $B$ factories, one place to look for the effects of new physics is in observables related to $B$ meson physics.  Therefore, in this paper we set up a parametrization for a set of observables associated with semileptonic $B$ decays into charmed mesons. Specifically, in Section~\ref{sec:def} we construct the low energy effective Lagrangian for the process $b\rightarrow c\ell{\bar {\nu}}_\ell$  induced by new physics.  We do so by including all the dimension six, four-fermion Lorentz scalars that contribute to the process, but excluding higher dimension operators.  We also introduce a set of real parameters bilinear in the coupling constants of this effective Lagrangian.  These parameters can be used to characterize the low energy effects of the new physics.  

Using the generalized interaction of Section~\ref{sec:def}, we then calculate at leading order in HQET~\cite{isgur,hqet,neubert} the contribution from new physics to the differential decay rates for  ${\bar B}\rightarrow D\ell{\bar {\nu}}_\ell$ and ${\bar B}\rightarrow D^*\ell{\bar {\nu}}_\ell$.  The results for these observables, expressed in the parametrization of Section~\ref{sec:def} appear in Section~\ref{sec:D} and Section~\ref{sec:D*} respectively.  The parametrization of Section~\ref{sec:def} also appears when we consider the corrections induced by physics beyond the Standard Model on the inclusive semileptonic decay of $B$ mesons into final states containing a $c$ quark, ${\bar B}\rightarrow X_c \ell{\bar {\nu}}_\ell.$  In Section~\ref{sec:Xc}, these deviations are calculated at leading order in perturbation theory, which is justified for $m_b\gg \Lambda_{\mbox{\tiny QCD}}$~\cite{ope}.  In Section~\ref{sec:disc} we use the results of the previous sections to discuss briefly the constraints which the existing experimental data imposes on the parametrization of Section~\ref{sec:def}.  

Finally, we note that new physics effects on observables related to $b\rightarrow c\ell{\bar\nu}_\ell$ have been studied extensively for restricted types of non-Standard model interactions.  The decays $B\rightarrow X\ell{\nu}_\ell$ (specially for decays into $\tau$-leptons) have been used to constrain classes of models with scalar interactions~\cite{kraw,kal,grzad1,hou,isidori,gross,coarasa}, such as those which arise in models with extended Higgs sectors.  Other constraints on such types of models have also come from observables related to the exclusive decays ${\bar B}\rightarrow D\ell{\nu}_\ell$~\cite{garisto,tanaka,kiers} as well as ${\bar B}\rightarrow D^*\ell{\nu}_\ell$~\cite{garisto,tanaka,grzad2}.  The process $b\rightarrow c\ell{\bar\nu}_\ell$ has also been used to analyze possible vector and axial vector couplings beyond the Standard Model $V-A$.  In particular, model-independent analyses of right-handed $b\rightarrow c$ quark current contributions to $B\rightarrow X\ell{\nu}_\ell$ and ${\bar B}\rightarrow D^*\ell{\nu}_\ell$ were performed in~\cite{voloshin,rizzo}.  These were motivated by an extension of the Standard Model, the Left-Right Symmetric model~\cite{gronau1}, which contains gauge bosons that couple to right-handed fermions.  Also inspired by this class of models were~\cite{garisto,gronau2}, who constructed observables based on the exclusive decays ${\bar B}\rightarrow D^{(*)}\ell{\nu}_\ell$ in order to constrain right-handed vector couplings.  The combined effects of non-Standard Model scalar, pseudoscalar, vector, and axial vector interactions were studied in~\cite{wu} within the context of $CP$ violating polarization observables related to ${\bar B}\rightarrow D\tau{\nu}_\tau$ and ${\bar B}\rightarrow D^*\ell{\nu}_\ell.$

\section{Definitions}
\label{sec:def}
Our starting point for the calculation of semileptonic $B$ decays to charmed mesons is the following interaction Hamiltonian:
\begin{equation}
\label{eq:int}
{\cal H}_{\mbox{\tiny int}}={\cal H}_{\mbox{\tiny{SM}}} + \left[{4 G_F \over\sqrt{2}}V_{cb}\sum_{\gamma,\mu,\epsilon} g^\gamma_{\mu\epsilon} [{\bar c}\Gamma^\gamma  b_\mu] [{\bar \ell}_\epsilon\Gamma^\gamma{\nu_\ell}] + \mbox{h.c.}\right].
\end{equation}
${\cal H}_{\mbox{\tiny{SM}}}$ is the effective $V-A$ Hamiltonian which mediates the Standard Model Weak interaction process $b\rightarrow c \ell{\bar \nu}_\ell$ for energy scales much less than $m_W$.  The second term represents the low-energy effective interaction generated by new physics at some energy higher than $m_W$.  Rather than adopt a particular model for this new physics, we include all the lowest dimension operators which contribute to $b\rightarrow c \ell{\bar \nu}_\ell$ at tree level and which also respect the Lorentz and gauge invariance of the Standard Model.  These are the dimension six four-fermi operators which appear in Eq.~(\ref{eq:int}).  One way to organize their contribution~\cite{tau} is to sum the index $\gamma$ over $S,V,T,$ with
\begin{eqnarray}
\Gamma^S=1; & (\Gamma^V)^\mu =\gamma^\mu; & (\Gamma^T)^{\mu\nu} =\sigma^{\mu\nu}\equiv\frac{i}{2}[\gamma^\mu,\gamma^\nu],
\end{eqnarray}
while projecting the $b$-quark and lepton $\ell$ into fields of definite chirality (respectively denoted by $b_\mu$ and $\ell_\epsilon$, with $\mu,\epsilon\in\{L,R\}$) and summing over all values of $\mu,$ $\epsilon$.  Given $\gamma,$ the chirality of the $c$ quark is fixed by $\mu$, and the chirality of the neutrino field $\nu_\ell$ is fixed by the value of $\epsilon.$  Note in particular that we are including operators which contain a right-handed neutrino field.  These occcur for $\gamma=S,T$ and $\epsilon=L,$ or $\gamma=V$ and $\epsilon=R$. 

The complex coefficients $g^\gamma_{\mu\epsilon}$ are a set of twelve dimensionless coupling constants (scaled by the Fermi constant $G_F$), one for each choice of $\gamma,$ $\mu,$ and $\epsilon.$  (In the abscence of right-handed neutrinos, $g^S_{\mu L}=g^V_{\mu R}=g^T_{\mu L}=0,$ and the number of coefficients would be reduced to six.)  If written in terms of the fields that transform as definite representations of the Standard Model $SU(3)\times SU(2)\times U(1)$ gauge group, no insertions of the Standard Model Higgs doublet are needed to make the dimension six operators of Eq.~(\ref{eq:int}) gauge invariant (we take the right-handed neutrino to be a Standard Model gauge singlet).  Dimensional analysis then implies that $g^\gamma_{\mu\epsilon}\sim{\mathcal O}(m_W^2/M^2),$ where $M$ is the energy scale of the new physics.  

Only certain real combinations of these coupling constants will appear in the expressions for the calculated observables.  To first order in $g^\gamma_{\mu\epsilon},$ we have
\begin{eqnarray}
\label{eq:1st}
\nonumber \beta_0 &=& 2\hbox{Re}\left\{g^V_{LL}\right\},\\
\nonumber \beta''_0 &=& 2 \hbox{Re}\left\{g^V_{RL}\right\},\\ 
\nonumber \epsilon_0 &=& 2 \hbox{Re}\left\{g^S_{LR}\right\},\\
\nonumber \epsilon'_0 &=& 2 \hbox{Re}\left\{g^S_{RR}\right\},\\
\rho_0 &=& 2 \hbox{Re}\left\{g^T_{LR}\right\},
\end{eqnarray}
which in principle could be present even in the abscence of a right-handed neutrino.  However, the last three of these can only appear for non-zero lepton mass.  We expect these five parameters, which arise by interference with ${\cal H}_{\mbox{\tiny{SM}}}$, to give the most significant contribution to signals of new physics.  Eq.~(\ref{eq:1st}) includes only the interference with the leading order part of the Standard Model term.  In reality, it should be modified by higher order Standard Model effects (such as radiative corrections), which we ignore here.  

The remaining parameters are second order in the  $g^\gamma_{\mu\epsilon}$ and are suppressed in comparison to the previous set.  We classify them by the following types:  
\begin{itemize}
\item Scalar-scalar:   
\begin{eqnarray}
\nonumber \alpha &=& \left|g^S_{LL}\right|^2 + \left|g^S_{LR}\right|^2  + \left|g^S_{RL}\right|^2  + \left|g^S_{RR}\right|^2, \\
\alpha'&=& 2 \hbox{Re}\left\{g^S_{RL} {g^S_{LL}}^\ast + g^S_{LR} {g^S_{RR}}^\ast\right\}.
\end{eqnarray}

\item Vector-vector:
\begin{eqnarray}
\nonumber \beta &=& \left|g^V_{LL}\right|^2 + \left|g^V_{RR}\right|^2, \\
\nonumber \beta'&=& \left|g^V_{LR}\right|^2  + \left|g^V_{RL}\right|^2,   \\
\beta''&=& 2 \hbox{Re}\left\{g^V_{RL} {g^V_{LL}}^\ast + g^V_{LR} {g^V_{RR}}^\ast\right\}.
\end{eqnarray}
\item Tensor-tensor:
\begin{equation}
\gamma = \left|g^T_{LR}\right|^2  + \left|g^T_{RL}\right|^2.
\end{equation}
\item Scalar-tensor
\begin{eqnarray}
\nonumber \delta &=& 2 \hbox{Re}\left\{g^T_{RL} {g^S_{RL}}^\ast + g^T_{LR} {g^S_{LR}}^\ast\right\},\\
\delta' &=& 2 \hbox{Re}\left\{g^T_{RL} {g^S_{LL}}^\ast + g^T_{LR} {g^S_{RR}}^\ast\right\}.
\end{eqnarray}
\item Scalar-vector (only appear for non-zero lepton mass):
\begin{eqnarray}
\nonumber \epsilon &=& 2 \hbox{Re}\left\{g^V_{LL} {g^S_{LR}}^\ast + g^V_{RL} {g^S_{RR}}^\ast+ g^V_{LR} {g^S_{LL}}^\ast + g^V_{RR} {g^S_{RL}}^\ast\right\},\\
\epsilon' &=& 2 \hbox{Re}\left\{g^V_{RL} {g^S_{LR}}^\ast + g^V_{LL} {g^S_{RR}}^\ast+ g^V_{RR} {g^S_{LL}}^\ast + g^V_{LR} {g^S_{RL}}^\ast\right\}.
\end{eqnarray}
\item Tensor-vector (only appear for non-zero lepton mass):
\begin{eqnarray}
\nonumber \rho &=& 2 \hbox{Re}\left\{g^V_{LL} {g^T_{LR}}^\ast + g^V_{RR} {g^T_{RL}}^\ast\right\},\\
\rho'&=&  2\hbox{Re}\left\{g^V_{RL} {g^T_{LR}}^\ast + g^V_{LR} {g^T_{RL}}^\ast\right\}.
\end{eqnarray}
\end{itemize}

\section{${\bar B}\rightarrow D \ell\bar{\nu}_\ell$}
\label{sec:D}
As in~\cite{neubert}, we write, in the rest frame of the $\bar B$-meson:
\begin{equation}
\label{eq:D}
{d^2 \Gamma\over dq^2 d(\cos\theta)} = {G_F^2 \left|V_{cb}\right|^2 \over 768\pi^3}\left|{\mathbf p}_D\right|{q^2-m_\ell^2 \over m_B^2}\left[(1+\cos\theta)^2\left|H_+\right|^2 + (1-\cos\theta)^2\left|H_-\right|^2 + 2 \sin^2\theta\left|H_0\right|^2\right],
\end{equation}
where ${\bf p}_D$ is the $D$ three-momentum in the $\bar B$ rest frame, $\theta$ is the angle between the lepton and the $D$ meson in the rest frame of the `virtual $W$-boson,' (i.e, the frame in which ${\mathbf p}_B = {\mathbf p}_D$)  and $q^2=(p_B-p_D)^2$. $\left|H_\pm\right|^2$ and $\left|H_0\right|^2$ are helicity amplitudes, which we decompose as:
\begin{eqnarray}
\nonumber \left|H_\pm\right|^2 &=& \left|H^{\mbox{\tiny (SM)}}_\pm\right|^2+\Delta^{(1)}_\pm + \Delta^{(2)}_\pm,\\
\left|H_0\right|^2 &=&\left|H^{\mbox{\tiny (SM)}}_0\right|^2 +\Delta^{(1)}_0+ \Delta^{(2)}_0.
\end{eqnarray}
In these expressions, $\left|H^{\mbox{\tiny (SM)}}_\pm\right|^2$ and $\left|H^{\mbox{\tiny (SM)}}_0\right|^2$ are the helicity amplitudes computed assuming only the Standard Model $V-A$ coupling.  The terms $\Delta^{(1)}_\pm$, $\Delta^{(1)}_0$ are contributions from the interference of the Standard Model part of the amplitude (taken only to leading order) and other terms in Eq.~(\ref{eq:int}).  Finally, the terms $\Delta^{(2)}_\pm$ and $\Delta^{(2)}_0$ are second order in the new interactions of Eq.~(\ref{eq:int}).  Here we calculate the interference and the second order terms by matching hadronic matrix elements of the quark-quark operators in Eq.~(\ref{eq:int}) to HQET matrix elements, neglecting both perturbative ${\mathcal O}(\alpha_s)$ and non-perturbative ${\mathcal O}(1/m_Q)$ corrections.  Higher order corrections to the Standard Model contribution to the decays ${\bar B}\rightarrow D^{(*)}\ell{\bar \nu}_\ell$ can be found in~\cite{ligeti}.  

Writing $w=(m_B^2 + m_D^2 -q^2)/2 m_B m_D,$ and denoting the Isgur-Wise function~\cite{isgur} by $\xi(w)$, we find for the interference tems:
\begin{eqnarray}
\nonumber \Delta^{(1)}_\pm&=& \frac{3 m_B m_D}{q^2} (1+w)\xi(w)^2\biggl[{m_\ell^2\over q^2}(\beta_0+\beta''_0)\left\{\left(w\pm\sqrt{w^2-1}\right)m_B^2\right.\\
\nonumber & &\hspace{0.5cm}\left. \mbox{}- 2 m_B m_D + \left(w\mp\sqrt{w^2-1}\right)m_D^2\right\}\\
\nonumber & &  \mbox{}+ \frac{m_\ell}{2}\epsilon_0\left\{(m_B - m_D) (1+w) \pm (m_B+m_D)\sqrt{w^2-1}\right\}\\
& &\mbox{}+ 2m_\ell\rho_0\left\{(m_B+m_D) (w-1) \pm (m_B-m_D)\sqrt{w^2-1}\right\}\biggr],
\end{eqnarray}
and
\begin{eqnarray}
\nonumber \Delta^{(1)}_0 &=& \frac{3 m_B m_D}{q^2}(1+w)\xi(w)^2\biggl[ (\beta_0+\beta''_0)\Bigl\{m_\ell^2+(m_B+m_D)^2(w-1)\Bigr\}\\
& &  \mbox{}+ {m_\ell\over 2} \epsilon_0 (1+w)(m_B-m_D) + 2 m_\ell \rho_0 (w-1) (m_B + m_D) \biggr].
\end{eqnarray}
Likewise, the second order terms are
\begin{eqnarray}
\nonumber \Delta^{(2)}_\pm&=&  {3\over 2} m_B m_D (1+w) \xi(w)^2\Bigl[(\alpha+\alpha')(1+w) + 16\gamma(w-1)\\
& &  \mbox{}\pm 4(\delta+\delta')\sqrt{w^2-1}\Bigr] + \Delta^m_\pm,\\
\nonumber \Delta^{(2)}_0&=& {3 m_B m_D\over 2 q^2}(1+w)\xi(w)^2\Bigl[(\alpha+\alpha')(1+w)q^2\\
& & \mbox{}+2\big\{(\beta+\beta'+\beta'')(m_B + m_D)^2 - 8q^2\gamma\big\}(w-1)\Bigl] + \Delta^m_0,
\end{eqnarray}
where $\Delta^m_\pm$ and  $\Delta^m_0$ represent corrections due to non-zero lepton mass (which are only non-negligible for decays into $\tau$-leptons):
\begin{eqnarray}
\nonumber \Delta^m_\pm &=& \frac{3 m_B m_D}{q^2}(1+w)\xi(w)^2 \biggl[\frac{m_\ell^2}{q^2}(\beta + \beta'+ \beta'')\left\{\left(w\pm\sqrt{w^2-1}\right)m_B^2\right.\\
\nonumber& &\left. \hspace{0.5cm} \mbox{} - 2 m_B m_D + \left(w\mp\sqrt{w^2-1}\right)m_D^2\right\}\\
\nonumber& &  \mbox{} +\frac{m_\ell}{2}\epsilon\left\{(m_B-m_D)(1+w) \pm (m_B + m_D) \sqrt{w^2-1}\right\}\\
& & \mbox{} + 2m_\ell\rho\left\{(m_B+m_D) (w-1) \pm (m_B-m_D)\sqrt{w^2-1}\right\}\biggr],\\
\nonumber \Delta^m_0 &=& \frac{3 m_B m_D}{q^2}(1+w)\xi(w)^2 \Bigl[m_\ell^2\{(\beta + \beta'+ \beta'') + 16\gamma (w-1) \}\\
& & \mbox{} + \frac{m_\ell}{2}\epsilon(1+w)(m_B-m_D) + 2m_\ell\rho (w-1)(m_B + m_D)\Bigr].
\end{eqnarray}

\section{${\bar B}\rightarrow D^* \ell\bar{\nu}_\ell$}
\label{sec:D*}
As in the previous case, we write in the $\bar B$ rest frame
\begin{equation}
\label{eq:D*}
{d^2 \Gamma\over dq^2 d(\cos\theta)} = {G_F^2 \left|V_{cb}\right|^2\over768\pi^3}\left|{\bf p}_{D^*}\right|{{q^2-m_\ell^2}\over m_B^2}\left[(1+\cos\theta)^2\left|H_+\right|^2 + (1-\cos\theta)^2\left|H_-\right|^2 + 2 \sin^2\theta\left|H_0\right|^2\right].
\end{equation}
The kinematic variables ${\bf p}_{D^*}$, $\theta$ and $q^2$ are defined as before (just replace $D$ with $D^*$ in the definitions of the previous section).  Also as before, we separate the helicity amplitudes as
\begin{eqnarray}
\nonumber \left|H_\pm\right|^2 &=& \left|H^{\mbox{\tiny (SM)}}_\pm\right|^2+\Delta^{(1)}_\pm + \Delta^{(2)}_\pm,\\
\left|H_0\right|^2 &=&\left|H^{\mbox{\tiny (SM)}}_0\right|^2 +\Delta^{(1)}_0+ \Delta^{(2)}_0.
\end{eqnarray}
Once again, the interference terms only take into account the leading order part of the Standard Model amplitude.  Calculating these as well as the second order terms by matching hadronic matrix elements onto HQET ones (neglecting ${\mathcal O}(\alpha_s)$ or ${\mathcal O}(1/m_Q)$ corrections), we find:  
\begin{eqnarray}
\nonumber \Delta^{(1)}_\pm &=& \frac{3 m_B m_{D^*}}{q^2}(1+w)\xi(w)^2 \biggl[2 (\beta_0 w -\beta''_0)q^2 + \frac{m_\ell^2}{q^2}(\beta_0-\beta''_0)\left\{\left(w\pm\sqrt{w^2-1}\right) m_B^2\right.\\ 
\nonumber& & \left.\hspace{0.5cm}\mbox{} - 2 m_B m_{D^*} + \left(w\mp\sqrt{w^2-1}\right)m_{D^*}^2\right\}\\
\nonumber& & \mbox{} + \frac{m_\ell}{2}(\epsilon_0'-\epsilon_0)\left\{(m_B + m_{D^*})(w-1)\pm(m_B-m_{D^*})\sqrt{w^2-1}\right\}\\
& &  \mbox{} - 2m_\ell\rho_0\left\{m_B(5+w)-m_{D^*}(1+5w)\pm (m_B+5 m_{D^*})\sqrt{w^2-1}\right\}\biggr],\\
\nonumber \Delta^{(1)}_0  &=& {3 m_B m_{D^*}\over q^2}(1+w)\xi(w)^2\Bigl[ (1+w)(m_B-m_{D^*})^2 (\beta_0-\beta''_0) \\
\nonumber & & \mbox{} + m_\ell^2\{\beta_0(2w-1)-\beta''_0\} + \frac{m_\ell}{2}(\epsilon'_0-\epsilon_0)(m_B+m_{D^*})(w-1)\\
& &  \mbox{} - 2m_\ell \rho_0\{m_B(5+w) - m_{D^*}(1+5w)\}\Bigr].
\end{eqnarray}
The second order terms are
\begin{eqnarray}
\nonumber \Delta^{(2)}_\pm &=& {3\over 2} m_B m_{D^*} (1+w) \xi(w)^2\Bigl[(\alpha-\alpha')(w-1) + 4\left\{(\beta+\beta')w-\beta''\right\}\\
& & \mbox{}+ 16\gamma(1+w) \pm4\left(\beta'-\beta + \delta-\delta'\right)\sqrt{w^2-1}\Bigr] + \Delta^m_\pm,\\
\nonumber \Delta^{(2)}_0 &=& {3 m_B m_{D^*}\over 2 q^2} (1+w) \xi(w)^2\Bigl[(\alpha-\alpha')(w-1)q^2\\
\nonumber& &  \mbox{} + 2(\beta+\beta'-\beta'') (m_B-m_{D^*})^2 (1+w)\\
& & \mbox{} + 16\gamma\left\{(m_B^2+m_{D^*}^2)(3w-1) + 2 m_B m_{D^*} (w^2 + w -4)\right\}\Bigr] +\Delta^m_0,
\end{eqnarray}
with the following corrections due to the effects of a non-zero lepton mass:
\begin{eqnarray}
\nonumber\Delta^m_\pm &=& \frac{3 m_B m_{D^*}}{q^2}(1+w)\xi(w)^2 \biggl[\frac{m_\ell^2}{q^2} (\beta+\beta'+32\gamma -\beta'')\left\{\left(w\pm\sqrt{w^2-1}\right) m_B^2\right. \\ 
\nonumber& & \hspace{0,5cm}\left. \mbox{} - 2 m_B m_{D^*}+\left(w\mp\sqrt{w^2-1}\right)m_{D^*}^2\right\}\\
\nonumber & &  \mbox{} +\frac{m_\ell}{2}(\epsilon'-\epsilon)\left\{(m_B + m_{D^*})(w-1)\pm(m_B-m_{D^*})\sqrt{w^2-1}\right\}\\
\nonumber & & \mbox{}+ 2m_\ell\left\{(\rho' m_B +\rho m_{D^*})(1+5w)\pm 5(\rho' m_B - \rho m_{D^*})\sqrt{w^2-1}\right.\\
& &\left. \hspace{0,5cm}\mbox{} - (\rho m_B +\rho' m_{D^*}) (5+w) \mp (\rho m_B - \rho' m_{D^*})\sqrt{w^2-1}\right\}\biggr],\\
\nonumber \Delta^m_0 &=&  \frac{3 m_B m_{D^*}}{q^2}(1+w)\xi(w)^2 \Bigl[m_\ell^2\{(\beta+\beta')(2w-1)  - \beta'' + 16\gamma(1+w)\}\\
\nonumber & & \mbox{} + \frac{m_\ell}{2}(\epsilon'-\epsilon)(m_B+m_{D^*})(w-1) + 2 m_\ell m_B\{\rho'(1+5 w) -\rho(5+w)\}\\
& & \mbox{}+ 2 m_\ell m_{D^*}\{\rho(1+5w) - \rho'(5+w)\}\Bigl].
\end{eqnarray}

\section{${\bar B}\rightarrow X \ell\bar{\nu}_\ell$}
\label{sec:Xc}

Starting from the operator product expansion (OPE), it can be shown that in the limit $m_b\gg \Lambda_{\mbox{\tiny QCD}}$ the inclusive semileptonic $B$ decay rate is equivalent to the perturbative quark level $b$ decay rate~\cite{ope}.  This makes it possible to calculate the effects of new physics on the inclusive decay ${\bar B}\rightarrow X_c \ell\bar{\nu}_\ell$ directly from Eq.~(\ref{eq:int}), by calculating the rate for the quark level process $b\rightarrow c\ell{\bar \nu}_\ell$.  We write the differential decay rate as
\begin{equation}
\label{eq:diff}
{d^3\Gamma\over {dq^2 dE_\ell dE_\nu}} = \left({d^3\Gamma\over {dq^2 dE_\ell dE_\nu}}\right)_{\mbox{\tiny (SM)}} + \left({d^3\Gamma\over {dq^2 dE_\ell dE_\nu}}\right)_{(1)} + \left({d^3\Gamma\over {dq^2 dE_\ell dE_\nu}}\right)_{(2)}, 
\end{equation}
where $q^2=(p_b-p_c)^2$ and $E_\ell$, $E_\nu$ are the $b$ quark rest frame energies of the lepton and neutrino respectively.  Eq.~(\ref{eq:diff}) is not meant to imply that the differential rate is a contribution from three distinct processes.  Instead, we are merely decomposing the observable into a piece arising solely from Standard Model physics, an interference term, and a term that is second order in the new physics.   At present, nonperturbative corrections to the Standard Model term are known to order $(\Lambda_{\mbox{\tiny QCD}}/m_b)^3$~\cite{gremm0},  while only the part proportional to $\alpha_s^2\beta_0$ of the NLO perturbative corrections is known~\cite{gremm1} ($\beta_0$ being the coefficient of the one-loop QCD beta function).  Here, we calculate the last two terms in Eq.~(\ref{eq:diff}), neglecting ${\mathcal O}(\alpha_s)$ or ${\mathcal O}(\Lambda_{\mbox{\tiny QCD}}/m_b)$ corrections.  Introducing the kinematic variables 
\begin{eqnarray}
\nonumber E_c &=&(m_b^2+m_c^2 - q^2)/2m_b,\\ 
\nonumber {\hat q}^2 &=& q^2/m_b^2,\\ 
{\hat E}_\nu &=& 2 E_\nu/m_b,
\end{eqnarray}
and $y=E_\ell/{E_\ell}_{\mbox{\tiny{max}}}$ with ${E_\ell}_{\mbox{\tiny{max}}}=(m_b^2-m_c^2+m_\ell^2)/2 m_b$, as well as $x=m_\ell/m_b$ and $z=m_c/m_b$, we find, in the rest frame of the $b$ quark
\begin{eqnarray}
\label{eq:diff1}
\nonumber \left({d^3\Gamma\over {dq^2 dE_\ell dE_\nu}}\right)_{(1)} &=& {G_F^2\left|V_{cb}\right|^2 m_b^2\over 32\pi^3}\delta\left(m_b-E_c-E_e-E_\nu\right)\biggl[-4z\beta''_0\left({\hat q}^2-x^2\right)\\
\nonumber & &   \mbox{} + 4\beta_0 {\hat E}_\nu\Bigl\{\left(x^2(1+y)+y(1-z^2)\right)-{\hat q}^2\Bigr\}\\
\nonumber & & \mbox{} + 2x\left((1-z^2)(1-y)-(1+y)x^2\right)\epsilon_0'\\
& &  +2{\hat E}_\nu xz(\epsilon_0-12\rho_0)\biggr],
\end{eqnarray}
which, as in the exclusive decays, includes only the interference with the leading order Standard Model contribution.  The remaining contributions are of second order in the new physics:
\begin{eqnarray}
\label{eq:diff2}
\nonumber \left({d^3\Gamma\over {dq^2 dE_\ell dE_\nu}}\right)_{(2)} &=& {G_F^2\left|V_{cb}\right|^2 m_b^2\over 32\pi^3}\delta\left(m_b-E_c-E_e-E_\nu\right)\biggl[\left({\hat q}^2-x^2\right)\Bigl\{\alpha(1+z^2-{\hat q}^2)\\
\nonumber & &\hspace{0.5cm} \mbox{} + 2z(\alpha'-2\beta'')\Bigr\}\\
\nonumber & &   \mbox{} + 4\beta {\hat E}_\nu\Bigl\{\left(x^2(1+y)+y(1-z^2)\right)-{\hat q}^2\Bigr\}\\
\nonumber & & \mbox{} + 2\left((1-z^2)(1-y)-(1+y)x^2\right)\Bigl\{2y(1+x^2-z^2)\beta'\\
\nonumber & & \hspace{0.5cm} \mbox{} + x(\epsilon'+12\rho')\Bigr\} \\
\nonumber & & \mbox{} -4\delta\Bigl\{{\hat E}_\nu\left(x^2(1+y)+y(1-z^2)\right)\\
\nonumber & & \hspace{0.5cm} \mbox{} + y(1+x^2-z^2)\left(x^2(1+y)-(1-z^2)(1-y)\right) -{\hat E}_\nu {\hat q}^2\Bigr\}\\
\nonumber & &  \mbox{} + 16\gamma \Bigl\{2{\hat E}_\nu\left(x^2(1+y)+y(1-z^2)\right) + \left(2y(1-y)(1-z^2)^2\right. \\
\nonumber & & \hspace{0.5cm} \left. \mbox{} + \left(1+z^2 -4(1-z^2)y^2\right)x^2-2y(1+y)x^4\right)\\
\nonumber & & \hspace{0.5cm}\mbox{} -{\hat q}^2(1+x^2+z^2+ 2{\hat E}_\nu)-{\hat q}^4\Bigr\}\\
& &  \mbox{}+2{\hat E}_\nu xz(\epsilon-12\rho)\biggr].
\end{eqnarray}
Integration of these expressions over $E_\nu$ is trivial.  Performing the $q^2$ integral over its physically allowed region gives the lepton spectrum
\begin{equation}
{d\Gamma\over dE_\ell} = \left({d\Gamma\over dE_\ell}\right)_{\mbox{\tiny (SM)}} + \left({d\Gamma\over dE_\ell}\right)_{(1)} + \left({d\Gamma\over dE_\ell}\right)_{(2)},
\end{equation}
where
\begin{eqnarray}
\label{eq:lep1}
\nonumber \left({d\Gamma\over dE_\ell}\right)_{(1)} &=& {{G_F^2 \left|V_{cb}\right|^2 m_b^4}\over 48\pi^3} F(x,y,z)\biggl[G_\beta (x,y,z)\beta_0\\
\nonumber & & \mbox{} - 3z\left(1-(1-z^2)y + (1-y)x^2\right) \left((1-z^2)y-(2-y)x^2\right)\beta''_0  \\
\nonumber & & \mbox{} + \frac{3}{2}xz\left(2-y(1-z^2)-yx^2\right)\left(1-(1-z^2)y + (1-y)x^2\right)(\epsilon_0 - 12\rho_0)\\
& & \mbox{} + 3x\left(1-(1-z^2)y + (1-y)x^2\right)^2\epsilon'_0 \biggr]
\end{eqnarray}
is the $q^2$ integral of Eq.~(\ref{eq:diff1}) and
\begin{eqnarray}
\label{eq:lep2}
\nonumber \left({d\Gamma\over dE_\ell}\right)_{(2)} &=&  {{G_F^2 \left|V_{cb}\right|^2 m_b^4}\over 48\pi^3} F(x,y,z)\biggl[G_\beta (x,y,z)(\beta+\frac{1}{4}\alpha)\\
\nonumber & & \mbox{} + \frac{3}{2}z\left(1-(1-z^2)y + (1-y)x^2\right)\left((1-z^2)y-(2-y)x^2\right)(\alpha'-2\beta'')\\
\nonumber & & \mbox{} + 3\left(1-(1-z^2)y + (1-y)x^2\right)^2(2y(1-z^2 + x^2)\beta'+x\epsilon'+12x\rho')\\
\nonumber & &  \mbox{} +\frac{3}{2}xz\left(2-y(1-z^2) - yx^2\right)\left(1-(1-z^2)y + (1-y)x^2\right)(\epsilon - 12\rho)\\
& & \mbox{} + G_\delta(x,y,z)\delta + 4G_\gamma(x,y,z)\gamma\biggr] 
\end{eqnarray}
is that of Eq.~(\ref{eq:diff2}).  In these last two expressions, the coefficient functions are defined by:
\begin{eqnarray}
F(x,y,z) &=& (1-y)^2 (1+ x^2 - z^2)^2 \sqrt{y^2(1-z^2+x^2)^2 - 4x^2}\over [1-(1-z^2)y + (1-y)x^2]^3,\\
\nonumber G_\beta (x,y,z) &=& y(1-z^2)\left\{3(1+z^2) -(5-4z^2-z^4)y + 2y^2(1-z^2)^2\right\}  \\
\nonumber & & - \left\{4(1+2z^2)-(13-4z^2-3z^4)y + 3(5-6z^2+z^4)y^2\right.\\
\nonumber & & \hspace{0.5cm}\left.  -6(1-z^2)^2 y^3\right\} x^2\\
\nonumber & & - \left\{4-13y +3(5-3z^2)-6(1-z^2)y^3\right\}x^4\\
& & +y\left\{3-5y+2y^2\right\}x^6,\\
\nonumber G_{\delta}(x,y,z) &=& y (1-z^2)^2 \left\{3-(7-z^2)y+4(1-z^2)y^2\right\} \\
\nonumber & & + \left\{4+8z^2+(5-8z^2+3z^4)y-3(7-10z^2+3z^4)y^2\right.\\
\nonumber & & \hspace{0.5cm}\left.+ 12(1-z^2)^2y^3\right\}x^2\\
\nonumber & & + \left\{4+(5-6z^2)y-3(7-5z^2)y^2+12(1-z^2)y^3\right\}x^4\\
& & +y\left\{3-7y+4y^2\right\}x^6,\\
\nonumber G_\gamma(x,y,z) &=& y(1-z^2) \left\{3(5+z^2)-(29-28z^2-z^4)y+14(1-z^2)^2 y^2\right\}\\
\nonumber & & - \left\{4(1+2z^2)-(49-28z^2-3z^4)y+3(29-38z^2+9z^4)y^2\right.\\
\nonumber & & \left.\hspace{0.5cm} -42(1-z^2)^2 y^3\right\}x^2\\
\nonumber & & - \left\{4-(49-12z^2)y + 3(29-19z^2)y^2-42(1-z^2)y^3\right\}x^4\\
& & +y\left\{15-29y+14y^2\right\}x^6.
\end{eqnarray}

\section{Discussion}
\label{sec:disc}

Any effects of physics beyond the Standard Model should become more apparent with the improved data from the next generation $B$ factories, some of which are scheduled to go on line in the near future.  Therefore, a detailed extraction of our parametrization of new physics should await these results.  For completeness, however, we will use the formulas derived in the previous sections together with the existing data to obtain bounds on some of the parameters of Section~\ref{sec:def}.  Our analysis is only meant to be illustrative and therefore should not be taken too seriously.

One place to look for constraints on the parametrization of Section~\ref{sec:def} is in the extraction of $|V_{cb}|$ from the exclusive decays ${\bar B}\rightarrow D^{(*)}e{\bar \nu}_e$.  Because in the zero recoil limit ($w\rightarrow 1$), the ${\mathcal O}(1/m_Q)$ HQET corrections to ${\bar B}\rightarrow D^{(*)}\ell{\bar \nu}_\ell$ vanish~\cite{luke}, and because $\xi(1)=1$~\cite{isgur}, these observables provide a theoretically clean way of extracting the CKM matrix element $|V_{cb}|$ from experiment~\cite{athanas,barish}.  In the abscence of new physics, we expect the value of $|V_{cb}|$ obtained from ${\bar B}\rightarrow D e{\bar \nu}_e$ to agree with that from the decay ${\bar B}\rightarrow D^* e{\bar \nu}_e$.  Therefore, a discrepancy between the two measurements of $|V_{cb}|$ could be used to put constraints on the parametrization of Section~\ref{sec:def}.  

Integrating Eq.~(\ref{eq:D}) and Eq.~(\ref{eq:D*}) over $\cos\theta$, and making a change of variables from $q^2$ to $w$, the exclusive decay rates can be written in the form~\cite{neubert}
\begin{eqnarray}
\nonumber {d\Gamma\over dw}({\bar B}\rightarrow D\ell{\bar \nu}_\ell) =  {G_F^2 |V_{cb}|^2 m_B^5\over 48\pi^3} (w^2-1)^{3/2} r^3 (1+r)^2 {\mathcal F}_D(w)^2,\\
\nonumber {d\Gamma\over dw}({\bar B}\rightarrow D^*\ell{\bar \nu}_\ell) = {G_F^2 |V_{cb}|^2 m_B^5\over 48\pi^3} (w^2-1)^{1/2}(w+1){r^*}^3 (1-r^*)^2\\
\times \left[1+{4w\over 1+w}{1-2r^*+{r^*}^2 \over (1-r^*)^2}\right] {\mathcal F}_{D^*}(w)^2,
\end{eqnarray}
where $r=m_D/m_B$ and $r^*=m_{D^*}/m_B$.  From the results of Section~\ref{sec:D} and Section~\ref{sec:D*}, we find
\begin{eqnarray}
\nonumber {\mathcal F}_D (w)^2 &=& {{\mathcal F}^{\mbox{\tiny (SM)}}_D (w)}^2 + (\beta_0 +\beta_0'')\xi(w)^2 + \biggl[\beta+\beta'+\beta''\\
\label{eq:FD}& & \mbox{} + {3(\alpha+\alpha')\over 2(1+r)^2}{w+1\over w-1}(1-2rw+r^2) +{8\gamma\over (1+r)^2}(1-2rw+r^2)\biggr]\xi(w)^2,\\
\nonumber {\mathcal F}_{D^*} (w)^2 &=& {{\mathcal F}^{\mbox{\tiny (SM)}}_{D^*} (w)}^2 + \left(\beta_0-{{(5+w){r^*}^2 -2 (1+5w)r^* + 5 +w}\over (1+5w){r^*}^2 -2(1+w+4w^2)r^* + 1 + 5w}\beta_0''\right)\xi(w)^2\\
\nonumber & & \mbox{} + \biggl[\beta + \beta' + \biggl\{{3\over 2}(w-1)(1-2wr^* +{r^*}^2)(\alpha-\alpha')\\
\nonumber & & \hspace{0.5cm}\mbox{}-\left((5+w){r^*}^2 -2 (1+5w)r^* + 5 +w\right)\beta''\\
\nonumber& &  \hspace{0.5cm}\mbox{}+8\left((1+5w){r^*}^2-2(4+w+w^2)r^* + 1+5w\right)\gamma\biggr\}\\
\label{eq:FD*}& & \times\left\{(1+5w){r^*}^2 -2(1+w+4w^2)r^* + 1 + 5w\right\}^{-1}\biggr]\xi(w)^2.
\end{eqnarray}
Note that although the decay rate for ${\bar B}\rightarrow D e{\bar \nu}_e$ is well-behaved at $w=1$, the quantity ${\mathcal F}_D (w)$ actually has a pole there.  This is only because the scalar contribution to the decay rate (the terms involving $\alpha$ and $\alpha'$) vanishes more slowly than $(w^2-1)^{3/2}$ as $w\rightarrow 1$.  In fact, the observation of this behavior as $w\rightarrow 1$ in the experimental data for ${\mathcal F}_D (w)$ could be seen as evidence for the possibility of scalar contributions to the process $b\rightarrow c\ell{\bar \nu}_\ell.$

We expect the second order parameters in Eq.~(\ref{eq:FD}) and Eq.~(\ref{eq:FD*}) to be suppressed by a factor of $(m_W/M)^2$ in comparison to the interference terms.  Ignoring their contribution, we find in the zero recoil limit:
\begin{equation}
\label{eq:ratio}
{{\mathcal F}_D (1)^2 \over {\mathcal F}_{D^*}(1)^2} \simeq\left({{{\mathcal F}^{\mbox{\tiny (SM)}}_D (1)}^2 + \beta_0 +\beta_0''\over {{\mathcal F}^{\mbox{\tiny (SM)}}_{D^*} (1)}^2 + \beta_0 - \beta_0''}\right).
\end{equation}
The CLEO Collaboration has used data on the exclusive decay ${\bar B}\rightarrow De{\bar \nu}_e$ at zero recoil to extract $|V_{cb}|{\mathcal F}_D (1)=0.0337\pm 0.044\pm 0.048^{+0.0053}_{-0.0012}$~\cite{athanas}.  From the  data on ${\bar B}\rightarrow D^*e{\bar \nu}_e$, it has also found $|V_{cb}|{\mathcal F}_{D^*} (1)=0.0351\pm 0.0019\pm 0.0018\pm 0.0008$~\cite{barish}.  If we neglect ${\mathcal O}(1/m_Q^2)$ corrections, ${{\mathcal F}_D^{\mbox{\tiny (SM)}}}(1)={{\mathcal F}_{D^*}^{\mbox{\tiny (SM)}}}(1)=1$ (the $1/m_Q$ corrections vanish automatically by Luke's theorem~\cite{luke}).  Eq.~(\ref{eq:ratio}) then reads
\begin{equation}
0.96\simeq {1+\beta_0+\beta_0''\over 1+\beta_0-\beta_0''},
\end{equation} 
which to lowest order in the parameters, gives $\beta_0''\simeq-0.02.$  

We can also use the data on the inclusive decays to put constraints on the parameters of Section~\ref{sec:def}.  The OPE for the lepton spectrum does not agree locally with the physical spectrum near the maximum lepton energy.  Therefore, the result of Section~\ref{sec:Xc} can only be compared with experiment by constructing an observable which integrates the OPE result over a sufficiently large region.  A suitable observable (introduced in~\cite{gremm2}) for comparison of the OPE with the inclusive data is given by
\begin{equation}
R_1 = {{\int_{1.5\mbox{\scriptsize GeV}} E_\ell {d\Gamma\over dE_\ell}dE_\ell}\over \int_{1.5\mbox{\scriptsize GeV}} {d\Gamma\over dE_\ell}dE_\ell},
\end{equation}
which is independent of the value of $|V_{cb}|$.  Using the inclusive $B\rightarrow Xe\bar{\nu}_e$ data from CLEO, the authors of~\cite{gremm2} extracted a central value of $R_1=1.7831\mbox{ GeV}.$  (This includes contributions from $b\rightarrow u e\bar{\nu}_e,$ which introduce an error of only a few percent.)  If we split $R_1$ into a piece coming from Standard Model physics alone, and a correction $\delta R_1$ due to new physics, we find 
\begin{equation}
\label{eq:dR1}
\delta R_1 \simeq {{\int_{1.5\mbox{\scriptsize GeV}} E_\ell {d\Gamma\over dE_\ell}dE_\ell}\over \int_{1.5\mbox{\scriptsize GeV}} \left({d\Gamma\over dE_\ell}\right)_{\mbox{\tiny SM}}dE_\ell}-{{R^{\mbox{\tiny (SM)}}_1\int_{1.5\mbox{\scriptsize GeV}}{d\Gamma\over dE_\ell} dE_\ell}\over \int_{1.5\mbox{\scriptsize GeV}} \left({d\Gamma\over dE_\ell}\right)_{\mbox{\tiny SM}}dE_\ell},
\end{equation}
where $R^{\mbox{\tiny (SM)}}_1$ is the Standard Model contribution to $R_1.$  For the purposes of this crude analysis, we can replace it by the full $R_1$ with negligible error.  Similar reasoning allows us to approximate the Standard Model part of the lepton spectrum by its leading order contribution.  Using Eq.~(\ref{eq:lep1}) and Eq.~(\ref{eq:lep2}) to do the integrals of Eq.~(\ref{eq:dR1}), 
\begin{eqnarray}
\label{eq:dR1n}
\nonumber \delta R_1 &\simeq& -0.001\alpha - 0.002\alpha' -0.005(\beta_0+\beta) -0.036\beta'\\
& & \mbox{} + 0.004(\beta_0''+\beta'')-0.031\delta-0.309\gamma,
\end{eqnarray}
where we have used the values $m_b=4.8\mbox{ GeV}$ and $m_c=1.4\mbox{ GeV}$~\cite{gremm2}.  Note, however, that the magnitude of the coefficients in Eq.~(\ref{eq:dR1n}) is rather sensitive to the particular values of $m_b$ and $z=m_c/m_b$.  Therefore, the bounds on new physics that are obtained from this equation will be highly dependent on the numerical values of the $b$ and $c$ quark masses that are used to evaluate them.

In~\cite{gremm2}, the Standard Model ${\mathcal O}(1/m_b^3)$ corrections to the theoretical value of $R_1$ are estimated to be
\begin{eqnarray}
\label{eq:dR1t}
\delta R_1 &=& -(0.4{\bar\Lambda}^3 + 5.5{\bar\Lambda}\lambda_1 + 6.8{\bar\Lambda}\lambda_2 + 7.7\rho_1)/{\bar m}_B^3,
\end{eqnarray}
where ${\bar m}_B=(m_B+3 m_{B^*})/4$ is the spin-averaged $B$ meson mass, which we take to be ${\bar m}_B=5.31\mbox{ GeV}$~\cite{pdb}.  Assume that the first three terms of this equation are fixed by the values ${\bar \Lambda}=0.39\mbox{ GeV},$ $\lambda_1=-0.19\mbox{ GeV}^2,$ $\lambda_2=0.12\mbox{ GeV}^2$ of~\cite{gremm2}.  The term involving $\rho_1$ has larger uncertainty.  Varying it between $\rho_1=0$ and the estimated value $\rho_1\simeq (300\mbox{ MeV})^3$ of~\cite{gremm2} gives 
\begin{equation}
\label{eq:range}
-8.5\times 10^{-4}\mbox{ GeV}\leq \delta R_1 \leq 5.3\times 10^{-4}\mbox{ GeV}.\end{equation}

Without a better knowledge of $\rho_1$, a contribution to $R_1$ from new physics in the range of Eq. (\ref{eq:range}) cannot be excluded, even if the other HQET parameters were known to arbitrary precision.  Therefore, assuming that the corrections to the theoretical value of $R_1$ from new physics also lie in this range, we can derive constraints on our parameters by comparing Eq.~(\ref{eq:range}) with Eq.~(\ref{eq:dR1}).  We will do so by choosing only one parameter from Eq.~(\ref{eq:dR1}) to be non-zero at a time.  Then the bounds are derived by taking the ratio of Eq.~(\ref{eq:range}) to the coefficient of that parameter in Eq.~(\ref{eq:dR1}).  Of course, this is not entirely satisfactory, since some of these parameters are not independent of each other.  For instance, setting $\alpha=0$ forces $\alpha'$ to vanish as well.  However, this procedure should be sufficient if all we are interested in is a rough numerical estimate.  Thus, given the uncertainty on $\rho_1$, we cannot rule out the following range for the scalar terms:
\begin{eqnarray}
0 \leq \alpha \leq 0.72; & -0.26 \leq\alpha' \leq 0.41,
\end{eqnarray}
Likewise, for the vector parameters 
\begin{eqnarray}
\nonumber -0.11\leq\beta_0 \leq0.18; & |\beta|<|\beta_0|,\\
\nonumber \begin{array}{c}0\leq \beta' \leq 0.023,\end{array}\\
-0.20\leq\beta_0'' \leq 0.13 ;&  |\beta''|<|\beta_0''|,
\end{eqnarray}
and finally $-0.017\leq\delta \leq 0.027$ and $0\leq\gamma\leq 0.003$.  Note in particular that the bounds on $\beta_0''$ are consistent with the value derived from the exclusive data.  

I would like to thank Mark Wise for guidance throughout the completion of this work.  Also, I would  like to thank Zoltan Ligeti for helpful comments on the manuscript.  This work was supported in part by the Department of Energy under grant number DE-FG03-92-ER 40701.

\end{document}